\documentclass{ws-procs9x6}

\begin{document}

\title{BARYONS AND LARGE $N_c$ IN HAPPY RESONANCE}

\author{R.F. LEBED$^*$}

\address{Department of Physics, Arizona State University,\\
Tempe, AZ 85287-1504, USA\\
$^*$E-mail: richard.lebed@asu.edu}

\begin{abstract}
I discuss recent developments in the large $N_c$ treatment of unstable
baryon resonances and the scattering amplitudes in which they appear.
These include pion photoproduction, extension to three-flavor
processes, decoupling of large $N_c$ artifacts, and combination of
this approach with results of chiral symmetry.
\end{abstract}


\bodymatter

\section{Introduction}\label{intro}

At the time of this writing, the 2006 {\it Review of Particle
Physics}\cite{PDG} has just landed on my desk.  Of this tome's 1232
pages, almost 10\% list properties of baryon resonances.  And yet, no
consistent picture has yet been developed that predicts their rich
spectroscopy with any degree of accuracy and consistency.  Quark
models accommodate a number of observed multiplets but predict
numerous others unsupported by experimental evidence, while treating
resonances as meson-baryon bound states explains some of the resonant
branching fractions but fails to reproduce the multiplet structure.
Both of these approaches derive from the underlying QCD theory only by
employing heuristic arguments that are often obscure.  And lattice
gauge theory, while not suffering this ambiguity, has far to develop
before it will be able to take on such an intricate tangle of unstable
states.

Since the beginning of 2002, Tom Cohen and I have been developing a
method\cite{CL1,CLcompat,CDLN1,CLpent,CDLN2,CLSU3,CDLM,CLpentSU3,CLSU3phenom,ItJt,CLdecouple,CLchiral}
that treats unstable baryon resonances consistently in the $1/N_c$
expansion of QCD as broad, unstable states [masses lying $O(N_c^0)$
above the ground-state baryons of mass $O(N_c^1)$, and widths of
$O(N_c^0)$].  The $1/N_c$ expansion is a well-defined
field-theoretical limit of QCD-like theories (specifically, QCD with
$N_c$ rather than 3 colors) and therefore escapes the criticisms of
the previous paragraph.  On the other hand, in its basic form the
$1/N_c$ expansion tells only how to count powers of $1/N_c$ and does
not by itself provide a means of computing dynamical quantities.  Even
so, it provides an organizing principle of remarkable
power,\cite{Prague} especially for baryons.

This formalism has been outlined in a number of earlier conference
proceedings,\cite{UMN,confs} the first of which was the preceding
installment of this CAQCD series.\cite{UMN} Since space here is
limited, I provide only a sketchy description of the origin of the
method in Sec.~\ref{old} and refer the interested reader to the prior
write-ups for more detail, and in Sec.~\ref{new} summarize findings
since the last CAQCD
conference.\cite{CLSU3,CDLM,CLpentSU3,CLSU3phenom,ItJt,CLdecouple,CLchiral}

\section{The Scattering Method} \label{old}

Chiral soliton models, particularly the Skyrme model, achieved a
degree of fashionability in the early to mid-1980's, owing to a series
of papers\cite{ANW} discussing their topological character and
connection to large $N_c$.  Prominent in this era were a number of
works noting the model independence of a number of their predictions,
particularly linear relations between meson-baryon scattering
amplitudes ({\it e.g.}, $S_{11} \! = \! S_{31}$).\cite{HEHW,Mattis} As
gradually became clear, these results are consequences of an
underlying symmetry imposed by the soliton's hedgehog configuration,
which is characterized by the quantum number $K$, where $\bf{K} \!
\equiv \! {\bf I} \! + \! {\bf J}$; the scattering amplitudes, labeled
by good $I$ and $J$ quantum numbers, are then obtained by forming
linear combinations of the $K$-labeled amplitudes in an exercise of
``Clebschology.''

Subsequently it was recognized that the underlying $K$ conservation
can be expressed, via crossing from the $s$ channel to the $t$
channel, in terms of the rule $I_t \! = \!  J_t$.\cite{Donohue,MM}
Nevertheless, the connection of these results to large $N_c$ depended
upon identifying the hedgehog configuration and its excitations with
this limit; it is certainly a reasonable approach, because the
ground-state baryons are built from the hedgehog, which exhibits the
maximal symmetry between $I$ and $J$ characterizing the ground-state
baryons.  Nevertheless, to see true compatibility with large $N_c$ one
needs a formalism that studies real meson-baryon scattering in the
large $N_c$ limit; this is provided by the consistency condition
approach developed in the early 1990's.\cite{DJM} One offshoot of this
program\cite{KapSavMan} showed the $I_t \! = \! J_t$ rule to be an
immediate consequence, and moreover that amplitudes with $|I_t \! - \!
J_t| \! = \! n$ are suppressed by at least $1/N_c^n$.  The pieces were
then in place to show that 1) the old linear amplitude relations based
on $K$ are true large $N_c$ results,\cite{CL1} 2) the relations apply
also to the baryon resonances embedded in these amplitudes,\cite{CL1}
and 3) $O(1/N_c^n)$ corrections can be incorporated\cite{CDLN2} by
including suppressed amplitudes with $|I_t \! - \! J_t| \! = \! n$.
Moreover, while the ground-state band of baryons, $I \! = \! J \! = \!
\frac 1 2, \frac 3 2, \ldots$, in the naive quark and chiral soliton
pictures for large $N_c$ is the same, the irreducible excited baryon
multiplets of the quark picture turn out to be reducible collections
of multiplets labeled by $K$ (``compatibility'').\cite{CLcompat} The
underlying $K$ conservation leads to a predictive pattern\cite{CL1} of
allowed and forbidden decays [explaining, {\it e.g.}, the dominance of
the $\eta N$ decay of $N(1535)$, since the $K \! \!  = \! 0$ pole in
the $I \! = \! J \! = \! \frac 1 2$ negative-parity channel couples
only to $\eta N$], tells what types of resonant multiplets are
allowed,\cite{CL1,CLcompat,CLpent} and respects the broadness of
resonances and configuration mixing between resonances of the same
quantum numbers.\cite{CDLN1}

\section{New Developments} \label{new}

\paragraph{Pion Photoproduction.}  The same approach that applies to
meson-baryon scattering may be employed as well for other cases of
physical interest.  As long as the quantum numbers and $1/N_c$
suppressions of the fields coupling to the baryon are known, the same
methods apply.  Here one has in mind such processes as
photoproduction,\cite{CDLM} electroproduction, or real or virtual
Compton scattering.  Photons, for example, carry both isovector and
isoscalar quantum numbers, the former dominating\cite{JJM} in baryon
couplings by a factor of $N_c$.  Amplitudes that include the leading
[relative $O(N_c^0)$] and first subleading [$O(1/N_c)$] isovector and
the leading [$O(1/N_c)$] isoscalar amplitudes then produces linear
relations among multipole amplitudes with relative $O(1/N_c^2)$
corrections.\cite{CDLM} While the relations obtained this way that
reflect the dominance of isovector over isoscalar amplitudes agree
quite impressively with data ({\it i.e.}, the amplitudes have the same
shape as functions of photon energy), a number of other relations,
particularly for magnetic multipoles, superficially appear to fare
badly.  However, in those cases the threshold behaviors still agree
quite well, followed by seemingly divergent behavior in the respective
resonant regions.  These discrepancies appear to be due to the
slightly different placement [at $O(\Lambda_{\rm QCD}/N_c) \! = \!
O(100 \, {\rm MeV})$] of resonances that are degenerate in the large
$N_c$ limit but with different $I,J$ values.  When a comparison of the
amplitude relations is performed by taking {\em on-resonance\/}
couplings\cite{PDG}, the linear relations good to $O(1/N_c^2)$ do
indeed produce agreement to within about 1 part in 9.\cite{CDLM}

\paragraph{Three Flavors.}  The work summarized in the previous
section all referred to systems either containing only $u$ and $d$
quarks, or those in which any heavier quarks are inert spectators.  In
nature, however, there appears to be some evidence that not only the
baryon ground-state band but resonances as well exhibit a degree of
SU(3) flavor symmetry.  The first task in such analysis in the $1/N_c$
expansion is to have the group theory under control, which requires
tables of SU(3) Clebsch-Gordan coefficients (CGC) as functions of
$N_c$; the standard ladder operator methods may be employed to derive
them.\cite{CLSU3} Since baryons at large $N_c$ have the quantum
numbers of $N_c$ quarks, their SU(3) representations are much larger
than for $N_c \! = \! 3$; for example, the analogue to the 3-color,
3-flavor octet is called ``{\bf 8}'': $(p,q) \! = \! [1, \frac 1 2
(N_c \! - \! 1)]$, which has $O(N_c^2)$ states.  One finds that the
linear amplitude relations connecting resonances of different $I$ and
$J$ for two flavors contain SU(3) CGC in the 3-flavor case, and
therefore link together members of distinct SU(3) multiplets with
various $J$ values.\cite{CLpentSU3} Such a phenomenon is known from
chiral soliton models, and occurs for the same reason: The group
theory is inherited from large $N_c$!  Once a single resonance is
found and its quantum numbers are measured, large $N_c$ tells what
other states of different $J$, $I$, and strangeness should be
degenerate to $O(\Lambda_{\rm QCD}/N_c) \! = \!  O(100 \, {\rm MeV})$
[not counting the additive $\sim$150~MeV contribution for each
$s$-quark] in both mass and width.  This is a very useful diagnostic
when one has a candidate exotic baryon, such as was the
$\Theta^+$.\cite{CLpent,CLpentSU3}

But it is also useful in identifying SU(3) partners of nonexotic
resonances.  For example, the $N(1535)$ should have strange
partners\cite{CLSU3phenom} that are also $\eta$-philic; and in fact,
there exists the $S_{01}$ state $\Lambda (1670)$ that lies only 5~MeV
above the $\eta \Lambda$ threshold (the phase space for $\pi \Sigma$
is 6 times larger), and yet its branching ratio to this channel is
10--25\%.  Other convincing examples\cite{CLSU3phenom} following the
large $N_c$ reasoning populate the sector of $\Lambda$ and $\Sigma$
resonances, but in many cases the uncertainty on branching ratios, or
even the existence of the resonances themselves, is questionable.  For
example, large $N_c$ makes definite statements about the spectroscopy
and decays of $\Xi$ and $\Omega$ resonances as well, but too little is
known about them experimentally to make definitive comparisons.

As noted in the previous section, the familiar quark model multiplets
at large $N_c$ such as the SU(6)$\times$O(3) (``{\bf 70}'',$1^-$)
actually form collections of distinct irreducible multiplets in large
$N_c$.  In the case just mentioned, one finds 5 such multiplets,
labeled by $K \! = \! 0, \frac 1 2, 1, \frac 3 2, 2$, whose masses can
differ at $O(N_c^0)$.  Those with $K \! = \! 0,1,2$ define the
multiplets with nonstrange members ({\it e.g.}, ``{\bf 8}'', ``{\bf
10}''), while those with $K \! = \! \frac 1 2 , \frac 3 2$ define
multiplets whose states of maximal hypercharge have a strange quark
({\it e.g.}, ``{\bf 1}'').

An unexpected result arises in the SU(3) group theory, in the form a a
theorem restricting which meson-baryon states have leading-order
[$O(N_c^0)$, by unitarity] CGC.\cite{CLpentSU3,CLSU3phenom} One can
prove that SU(3) CGC for {\em baryon}$_1$ + {\em meson}
$\leftrightarrow$ {\em baryon}$_2$ can be $O(N_c^0)$ only if $Y_{\rm
meson} \! = \!  Y_{B_2 , \, \rm max} \! - \! Y_{B_1 , \,
\rm max}$, which is to say that the meson must have a hypercharge
equal to the difference of the tops of the two baryon representations.
In particular, the dominant two-body decay mode of a $\Lambda$
resonance in an ``{\bf 8}'' is predicted to be $\pi \, \Sigma (1192)$,
while one in a ``{\bf 1}'' prefers $\overline{K} N$.  Evidence for
this remarkable result appears in, {\it e.g.}, $\Lambda (1520)$
$D_{03}$, where the coupling constant to $\overline{K} N$, once the
threshold $p^{2L+1} \! \to \! p^5$ factor is taken into account, is
4--5 [= $O(N_c^1)$] times larger than that for $\pi \Sigma$.  Indeed,
the SU(3) content of $\Lambda (1520)$ is traditionally assigned to be
dominantly singlet.

The SU(3) theorem is also the key ingredient in the proof that the
$I_t \! = \! J_t$ rule holds for three-flavor as well as two-flavor
processes, and moreover that processes with strangeness exchange are
suppressed (the $Y_t \! = \! 0$ rule).\cite{ItJt} This result
predicts, for example, that the process $K^- p \! \to \! \pi^+
\Sigma^-$ is suppressed in cross section by $1/N_c$ compared to $K^- p
\! \to \!  K^- p$.

\paragraph{Decoupling Spurious States.}

Since, as noted, large $N_c$ baryons inhabit much larger SU(3)
representations and also allow much higher spins than for $N_c \! =
\! 3$, most of these states must be $N_c \! > \! 3$ artifacts and need
to be decoupled from the theory as spurious if they appear in
amplitudes for physical $N_c \! = \! 3$ processes.  The effects of
such states must disappear smoothly and exactly at the value of $N_c$
where they become spurious.\cite{CLdecouple} It would be extremely
coincidental for such decoupling to occur through cancellation among
dynamical quantities, which depend sensitively upon nonperturbative
effects, not least of which are the precise values of quark masses.
We therefore argue\cite{CLdecouple} that the only sources of true
decoupling are group-theoretical in nature, either because the states
of interest have isospin or strangeness values too high to reach in
conventional $N_c \! = \! 3$ meson-baryon scattering, or lie in states
in SU(3) multiplets that decouple when $N_c \! = \! 3$.  Such
decoupling appears through factors of $1 \! - \! 3/N_c$, a very
special type of $1/N_c$ correction.  For example, for $N_c \! > \! 3$
the ``{\bf 1}'' contains states with $\Xi$ quantum numbers, and the
SU(3) coupling for $\overline{K} \Sigma$ to this state indeed contains
a factor of $\sqrt{1 \! - \! 3/N_c}$.

\paragraph{Interplay of Chiral and Large $N_c$ Limits.}

Since meson-baryon scattering at threshold has also been studied in
the context of another well-known expansion of strong interaction
physics, the chiral expansion, it is natural to ask how the chiral and
large $N_c$ limits cooperate and compete.\cite{CLchiral} The large
$N_c$ counting constraints for scattering amplitudes formally hold at
all energies, while the chiral results become increasingly better the
closer one approaches threshold.  Among these results are the famed
Weinberg-Tomozawa relation (between $I \! = \! 0,2$ $\pi N$ scattering
lengths).  Other well-known results relevant to the chiral $\pi$ in
meson-baryon scattering are the Adler-Weisberger (AW) and
Goldberger-Oehme-Miyazawa (GMO) cross section sum rules.

From the point of view of the $1/N_c$ expansion, scattering lengths
are simply the derivatives of scattering amplitudes at zero meson
energy.  One finds that the combination corresponding to the WT and
other chiral-limit relations are $O(N_c^0)$ but very well satisfied
experimentally (as expected for threshold results), while most
combinations that are $O(1/N_c)$ are not as well satisfied, but still
have magnitudes consistent with the naive counting [that is, a
dimensionless $O(1/N_c)$ combination is not larger than, say, 2/3].

The AW and GMO sum rules are not chiral limit results (They sum over
all allowed energies), but nevertheless refer to chiral $\pi$'s.  In
these cases one encounters the effect of the noncommutativity of the
two limits: In particular, both fail if the large $N_c$ limit is taken
prior to the chiral limit, a consequence of the fact that the
$\Delta$-$N$ mass difference is $O(1/N_c)$ and therefore falls below
$m_\pi$ for sufficiently large $N_c$. In that case the $\Delta$
becomes stable and must be treated on the same footing as the $N$.

\section{Conclusions}\label{concl}

The $1/N_c$ expansion provides a formalism in which to analyze
meson-baryon scattering processes, including their rich spectrum of
embedded resonances.  The resonances appear in multiplets that bear a
similarity to, but are more fundamental than, the old
SU(6)$\times$O(3) quark model multiplets.  These multiplets have
distinct dominant decay channels, {it e.g.}, by preferring
$\overline{K} N$ over $\pi \Lambda$ or $\pi \Sigma$.  They are
constrained by preferring $I_t \! = \!  J_t$, $Y_t \! = \! 0$
processes.  The problem of removing spurious ($N_c \! > \! 3$) states
from the theory has been solved, and the nature of the interplay
between the chiral and large $N_c$ limits in meson-baryon scattering
has been explored.

A number of interesting formal and phenomenological issues remain to
be resolved.  For example, all results presented here hold for
single-meson scattering; of course, constraints also occur for
scattering with, {\it e.g.}, $\pi \pi N$ final states, but these are
as yet unexplored.  Moreover, the three-flavor scattering results
presented thus far only scratch the surface of the detailed
phenomenological analysis that is now possible.  A reliable means now
exists to shed light on one of the darker corners of particle physics.

\section*{Acknowledgments}
I thank the organizers for their kind invitation to this interesting
and productive meeting.  This work was supported in part by the NSF
under Grants No.\ PHY-0140362 and PHY-0456520.

\end{document}